\documentclass[lettersize,journal]{IEEEtran}
\usepackage{amsmath,amsfonts}
\usepackage{algorithmic}
\usepackage{algorithm}
\usepackage{array}
\usepackage[caption=false,font=normalsize,labelfont=sf,textfont=sf]{subfig}
\usepackage{textcomp}
\usepackage{stfloats}
\usepackage{url} 
\usepackage{verbatim}
\usepackage{graphicx}
\usepackage{cite}
\usepackage{hyperref}
\usepackage{color}
\begin{document}

\title{Efficient Sensing Parameter Estimation with Direct Clutter Mitigation in Perceptive Mobile Networks}

\author{Hang Li, Hongming Yang, Qinghua Guo,~\IEEEmembership{Senior Member,~IEEE,} J. Andrew Zhang,~\IEEEmembership{Senior Member,~IEEE,}\\ Yang Xiang and Yashan Pang
\thanks{This work is supported by Zhejiang Provincial Natural Science Foundation (Grant LY22F010003), National Natural Science Foundation (Grant 62071163) and Project of Ministry of Science and Technology (Grant D20011). {\itshape (Corresponding author: Qinghua Guo)}}
\thanks{Hang Li, Hongming Yang and Yang Xiang are with the School of Electronics and Information, Hangzhou Dianzi University, Hangzhou 310018, China, (e-mails: \href{mailto:hangli@hdu.edu.cn}{hangli@hdu.edu.cn}; \href{mailto:222040342@hdu.edu.cn}{222040342@hdu.edu.cn}; \href{mailto:xiangyang0831@gmail.com}{xiangyang0831@gmail.com}).}
\thanks{Qinghua Guo is with the School of Electrical, Computer and Telecommunications Engineering, University of Wollongong, Wollongong, NSW 2522, Australia, (e-mail: \href{mailto:qguo@uow.edu.au}{qguo@uow.edu.au}).}
\thanks{J. Andrew Zhang is with the Global Big Data Technologies Centre, University of Technology Sydney, Ultimo, NSW 2007, Australia, (e-mail: \href{mailto:andrew.zhang@uts.edu.au}{andrew.zhang@uts.edu.au}).}
\thanks{Yashan Pang and Hang Li are with National Key Laboratory of Advanced Communication Networks, Academy of Network \& Communications of CETC, Hebei Shijiazhuang 050081, China, (e-mail: \href{mailto:yashanpang@163.com}{yashanpang@163.com}).}
}

\markboth{Journal of \LaTeX\ Class Files,~Vol.~14, No.~8, August~2021}%
{Shell \MakeLowercase{\textit{et al.}}: A Sample Article Using IEEEtran.cls for IEEE Journals}

\maketitle

\begin{abstract}
In this work, we investigate sensing parameter estimation in the presence of clutter in perceptive mobile networks (PMNs) that integrate radar sensing into mobile communications. Performing clutter suppression before sensing parameter estimation is generally desirable as the number of sensing parameters can be significantly reduced. However, existing methods require high-complexity clutter mitigation and sensing parameter estimation, where clutter is firstly identified and then removed. In this correspondence, we propose a much simpler but more effective method by incorporating a clutter cancellation mechanism in formulating a sparse signal model for sensing parameter estimation. In particular, clutter mitigation is performed directly on the received signals and the unitary approximate message passing (UAMP) is leveraged to exploit the common support for sensing parameter estimation in the formulated sparse signal recovery problem. Simulation results show that, compared to state-of-the-art methods, the proposed method delivers significantly better performance while with substantially reduced complexity.

\end{abstract}


\section{Introduction}
As mobile communication technology advances from the fifth generation (5G) to the forthcoming sixth generation (6G), it embodies the relentless pursuit of higher data rates, lower latency, and expanded connectivity\cite{9144301}. The current trend is moving towards perceptive mobile networks (PMNs)\cite{9585321}, which not only support communication but also possess the capability to sense the surrounding environment\cite{8827589}. This integration of sensing and communication, often referred to as integrated sensing and communication (ISAC)\cite{10418473}, is poised to transform industries by facilitating applications such as autonomous driving, smart city and enhanced internet of things (IoT)\cite{9540344}. Implementing PMNs with ISAC requires advanced sensing techniques using communication signals \cite{10049809}. One of the important issues is accurate sensing parameter estimation in the presence of clutter \cite{9585321}. Clutter, namely the unwanted echoes from objects in the environment, can significantly degrade PMNs' performance by masking expected targets and interfering with signal transmission. It primarily refers to multipath signals from permanent or long-period static objects, which can considerably increase the number of sensing parameters. Therefore, it is desirable to perform clutter suppression before parameter estimation.

Some research works regarding clutter suppression applied to mobile networks have been reported in \cite{7944212, 8827589, 9203851}. A maximum-likelihood based amplitude estimation for clutter estimation in ISAC was developed in \cite{7944212}, where a simple clutter model was employed for sample averaging evaluation. The use of complex clutter models may result in higher residuals after clutter cancellation, deteriorating sensing performance. 
Leveraging the stability of signals across coherence time periods, a recursive moving averaging (RMA) based method was proposed to suppress clutter, where signals over a window are recursively averaged and smoothed with a forgetting factor \cite{8827589}. Although this method effectively reduces clutter from static paths by filtering signals at fixed intervals, the random and irregular demodulation reference signals (DMRS) used in 5G-NR in the time domain make it difficult to apply to PMNs. A Gaussian mixture model (GMM) and expectation maximization (EM) based clutter estimation (CE) and suppression method called GMM-EM-CE was proposed for PMNs in \cite{9203851}, where it works by statistically modeling channel states with Gaussian distributions \cite{7336572}, allowing differentiation between static clutter and dynamic signals. However, unknown number of multipath components in real environments cause challenges in determining the number of Gaussian mixture components in this method. In addition, it also requires matrix inversion in each cycle, leading to high computational complexity. 
These traditional methods require channel state information for clutter suppression and subsequently perform sensing parameter estimation, hence their performance is subject to errors due to inaccurate channel estimation.

In this correspondence, to achieve accurate sensing in the presence of clutter in PMNs, we integrate a clutter cancellation mechnisam inspired by multipulse canceller into sensing parameter estimation. This enables the proposed method to directly remove clutter with low complexity using the received signals. This contrasts with existing methods, such as GMM-EM-CE implemented with the high-complexity EM algorithm, where the clutter is firstly identified and then removed.  The problem of sensing parameter estimation with clutter mitigation is then formulated as a sparse signal recovery one with multiple measurement vectors (MMV). We then develop an efficient downlink direct sensing parameter estimation method leveraging the
unitary approximate message passing (UAMP) algorithm  \cite{9547768,DBLP:journals/corr/GuoX15}, where the common support of the sparse vectors is exploited and parameter association is performed. In particular, we exploit the periodic scanning property of synchronization signal blocks (SSBs) in 5G NR. The proposed method achieves omnidirectional sensing with the received SSBs, avoiding the limited directional sensing due to the randomness and irregularity of DMRS in the time domain and the need for initial channel estimation. It is much simpler but delivers enhanced performance. Simulation results show that our sensing parameter estimation method with direct clutter mitigation achieves significantly better performance compared to state-of-the-art methods. 
Notations: We use $(\cdot)^H$, $(\cdot)^T$ and $(\cdot)^\ast$ to denote the Hermitian transpose, transpose, and conjugate of a matrix/vector, respectively. The notation $|\cdot|$ denotes the element-wise absolute value and $\Vert \cdot \Vert_2$ the 2-norm of the matrix/vector. $\mathbf{A}(n,m)$ represents the element of the $n$th row and $m$th column in matrix $\mathbf{A}$, $\mathbf{a}(n)$ the $n$th element of $a$ and $\angle(\cdot)$ the phase of a complex number.

\begin{figure}[t]
    \centering
    \includegraphics[width=3.33in]{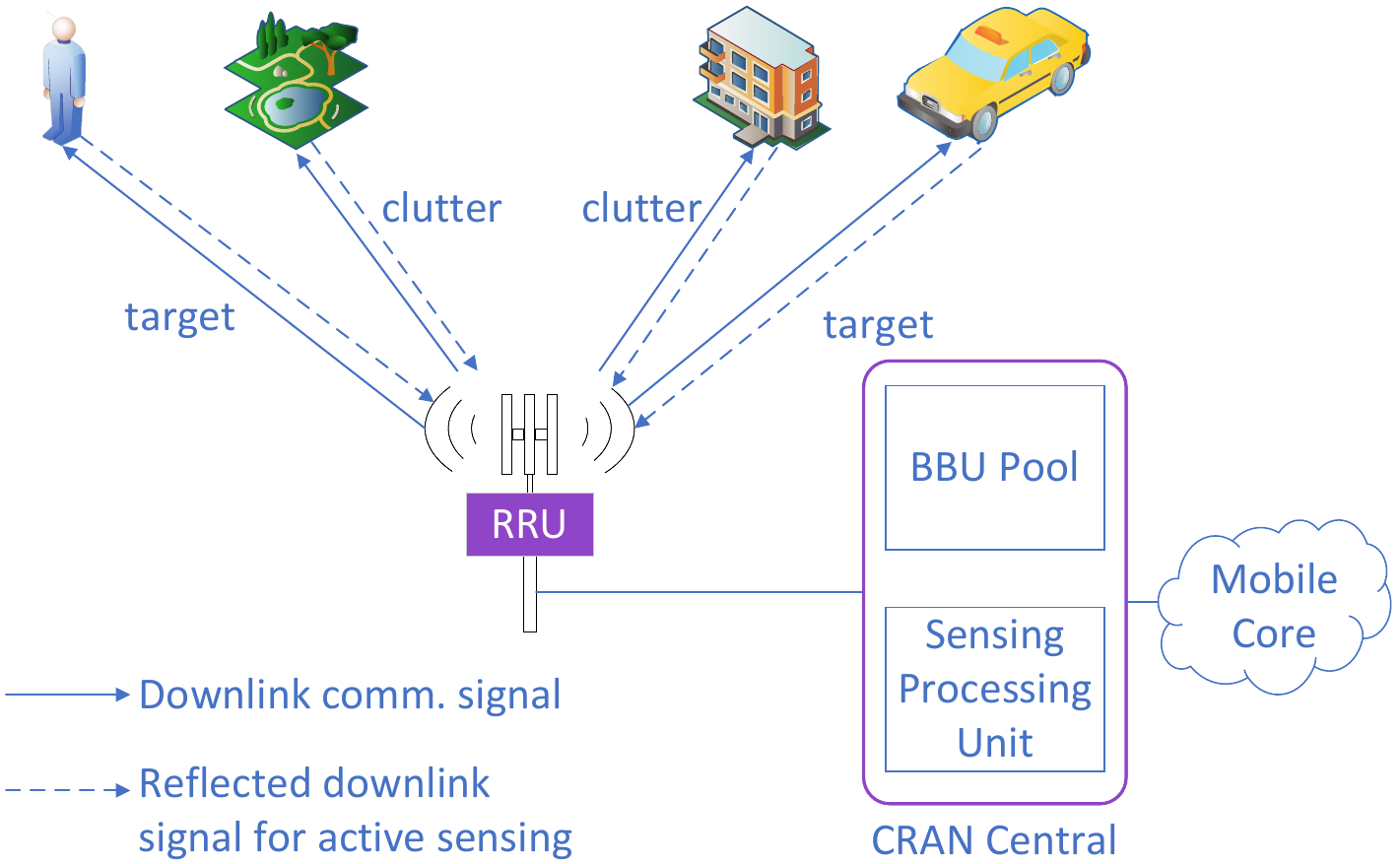}
    \caption{Illustration of the downlink active sensing.}\label{Model}
\end{figure}

\section{System and Signal Models}
As shown in Fig.~\ref{Model}, we consider a cloud-radio-access network (CRAN) based PMN~\cite{8827589}, where the remote radio unit (RRU) equipped with a uniform linear array (ULA) of \(M\) antennas performs downlink active sensing for estimating environmental parameters, i.e., using the echoes of its transmitted signals for sensing. Defining \(N\) as the aggregate number of subcarriers and \(B\) as the bandwidth, we have the subcarrier interval of \(\Delta f = \frac{B}{N}\), and the OFDM symbol period of \(T_s = \frac{N}{B} + T_p\), where \(T_p\) represents the cyclic prefix period. Assume a planar wave-front signal propagation, and the array response vector for the RRU can be represented as
\begin{align}
\mathbf{a}(M, \theta)=\left[1, e^{j \pi \sin (\theta)}, \ldots, e^{j \pi(M-1) \sin (\theta)}\right]^{H},
\end{align}
where $\theta$ is either an angle-of-departure (AoD) or angle-of-arrival (AoA), and the antenna spacing is assumed to be half wavelength. It is noted that, for downlink active sensing, the AoD is the same as the AoA for a target. We denote the AoDs (AoAs) as $\theta_{\ell}$, $\ell \in[1, L]$, where $L$ is the number of paths.

The $M \times M$ frequency domain channel matrix corresponding to the $n$th subcarrier of the $t$th OFDM symbol can be expressed as
\begin{align}
\mathbf{H}_{n,t}\!=\!\sum_{\ell=1}^{L}\!b_{\ell} e^{-j 2 \pi n \tau_{\ell} \Delta f} e^{j 2 \pi t f_{D, \ell} T_{s}} \mathbf{a}\left(M, \theta_{\ell}\right)\!\mathbf{a}^{T}\!\left(M,\theta_{\ell}\right),
\end{align}
where $b_{\ell}$ is the complex amplitude for the $\ell$th path, $\tau_{\ell}$ is the propagation delay, and $f_{D,\ell}$ is the corresponding Doppler frequency. The parameters $\{\tau_{\ell}, f_{D,\ell}, \theta_{\ell}, b_{\ell}\}$ are to be estimated, which can be used to identify targets' locations and speeds. We assume that the wireless channel keeps static within several milliseconds, during which these sensing parameters remain unchanged, and the Doppler phase over the samples in one OFDM symbol is approximately constant\cite{8827589}. 

To achieve omnidirectional sensing, we use the SSB signals of 5G NR \cite{3gpp.38.211} for clutter suppression and sensing in PMNs. The SSB includes the primary synchronization signal (PSS), the secondary synchronization signal (SSS) and the physical broadcast channel (PBCH). The PSS and SSS are used for synchronization and determination of the cell ID. The PBCH carries the master information block (MIB) including system configuration information, e.g., subcarrier spacing and frame structure. During an SSB scanning period that occupies part of the time resources in a half frame, the RRU transmits an SSB burst set consisting of multiple SSBs in different beamforming directions, achieving the full coverage of the cell. Each SSB spans 4 consecutive OFDM symbols and 20 resource blocks (RBs) in the frequency domain, each RB including 12 consecutive subcarriers. To exploit as many subcarriers as possible, we select the last three OFDM symbols of SSB for sensing. Denote the transmitted beamforming signals at the $n$th subcarrier of the $k$th $(k = 1, 2, 3)$ OFDM symbol within the $g$th $(g =1, 2, ..., G)$ SSB as $\mathbf{x}_{n,t} = \mathbf{w}_t s_{n,t},$
where $\mathbf{w}_t$ is the beamforming vector of the $t$th OFDM symbol $(t =4g + k)$, and $s_{n,t}$ is the associated phase-modulated signal with $|s_{n,t}|^2 = 1.$ The received signal of the RRU at the $n$th subcarrier and the $t$th OFDM symbol can be expressed as
\begin{equation}
    \begin{aligned}
        \mathbf{y}_{n,t}= & \sum_{\ell=1}^{L}b_{\ell}e^{-j2\pi n \tau_{\ell}\Delta f}e^{j2\pi tf_{D,\ell}T_s} \\
                          & \times\mathbf{a}(M,\theta_{\ell})\mathbf{a}^T(M,\theta_{\ell})\mathbf{x}_{n,t}+\mathbf{z}_{n,t}, \label{y_nt}
    \end{aligned}
\end{equation}
where $\mathbf{x}_{n,t}$ and $\mathbf{z}_{n,t}$ represent the transmitted signal vector and the noise vector, respectively.

\section{Sensing Parameter Estimation with Direct Clutter Mitigation}
\subsection{Signal Model with Direct Clutter Mitigation}

Inspired by the moving target indicator (MTI) technique in radar, we deal with the clutter by directly processing the received signal. This contrasts with the clutter suppression techniques in the literature where channel state information is required before clutter mitigation. In particular, leveraging the multipulse canceller technique \cite{9034521}, 
we exploit delays of SSB signal repetition to eliminate the components of echo signals received from permanent or long-period stationary targets in PMNs. We represent the received OFDM symbols with identical indices from $P+1$ successive SSB burst sets as $\mathbf{y}_{n,t}$, $\ldots,$ $\mathbf{y}_{n,t-PN_{s}}$, where $N_s$ is the number of OFDM symbols between two consecutive SSB burst sets. Following the multipulse cancellation principle, we directly process the received signals and have
\begin{align}
\mathbf{\tilde{y}}_{n,t}=&\sum_{p=0}^P(-1)^p\binom{P}{p}\mathbf{y}_{n,t-pN_s}\nonumber \\ 
\overset{(a)}{=}& \sum_{\ell =1}^L{a_{\ell} b_{\ell}e^{-j2\pi n\tau _{\ell}\Delta f}e^{j2\pi tf_{D,l}T_s}}\nonumber \\
&\times \mathbf{a}\left( M,\theta _{\ell} \right) \mathbf{a}^T \left( M,\theta _{\ell} \right)\mathbf{x}_{n,t} + \mathbf{\tilde{z}}_{n,t},
\label{ynttilde}
\end{align}
where $(a)$ holds because $\mathbf{x}_{n,t}$, $\ldots,$ $\mathbf{x}_{n,t-PN_{s}}$ are identical with the same beamforming for SSB, $\binom{P}{p}=\frac{P!}{p!(P-p)!}$ and $\mathbf{\tilde{z}}_{n,t}=\sum_{p=0}^P(-1)^p\binom{P}{p}\mathbf{z}_{n,t-pN_{s}}$. 
\begin{align}
a_{\ell} =& \sum_{p=0}^P(-1)^p\binom{P}{p}e^{-j2\pi f_{D,l}pN_sT_s}
\nonumber\\
=&(2j\sin{(\pi f_{D,l}N_sT_s}))^Pe^{-j\pi Pf_{D,l}N_sT_s} \label{al}
\end{align}
is the frequency response of the clutter canceller.
 It can be easily verified from (\ref{al}) that, when $f_{D,\ell}=0$, $a_\ell=0$ and thus the clutter can be eliminated from (\ref{ynttilde}). To deal with clutter with near-zero Doppler-frequencies, i.e., $f_{D,l} \approx 0$, a larger $P$ may be preferred, as the clutter canceller has deeper notches\cite{mahafza2003matlab}.  This will significantly reduce the number of parameters to be sensed and enhance the sensing accuracy of dynamic multipath or multipath of interests with nonzero Doppler frequencies.

Multiplying (\ref{ynttilde}) by $s_{n,t}^\ast$, we have a data-free signal model as
\begin{equation}
    \begin{aligned}            
    \mathbf{r}_{n,t} =& s_{n,t}^\ast \mathbf{\tilde{y}}_{n,t}= \sum_{\ell=1}^{L}  e^{-j2\pi n \tau_{\ell}\Delta f}\mathbf{b}_{\ell,t} + \mathbf{z}^\prime_{n,t},\label{rnt}
    \end{aligned}
\end{equation}
where $\mathbf{b}_{\ell,t} = a_{\ell} b_{\ell} e^{j2\pi tf_{D,\ell}T_s}  \mathbf{a}(M,\theta_{\ell}) \mathbf{a}^T(M,\theta_{\ell}) \mathbf{w}_t$ contains all sensing parameters but delay, and $\mathbf{z}^\prime_{n,t}$ is the processed noise vector. Then concatenating $\mathbf{r}_{n,t}$ for all $N$ subcarriers, we have
\begin{equation}
    \mathbf{R}_t = [\mathbf{r}_{1,t}, \ldots, \mathbf{r}_{N,t}]^T = \mathbf{C} \mathbf{B}_t + \mathbf{Z}^\prime_t,\label{R_t}
\end{equation}
where $\mathbf{C} = [\mathbf{c}(\tau_1), ..., \mathbf{c}(\tau_L)]$ includes the delays of all multipaths with $\mathbf{c}(\tau_{\ell}) = [e^{-j2\pi \tau_{\ell}\Delta f}, ..., e^{-j2\pi N\tau_{\ell}\Delta f}]^T$, and $    \mathbf{B}_t = [\mathbf{b}_{1,t}, \ldots, \mathbf{b}_{L,t}]^T$.

To facilitate delay estimation, we quantize the continuous delay term $e^{-j2 \pi n\tau_{\ell}\Delta f}$ to its nearest grid $e^{-j2 \pi n\ell^\prime/N_d}$, where $\ell^\prime/N_d$ denotes quantized $\tau_{\ell}\Delta f$ and the grid size $N_d \gg L$. Note that, a wideband system can generally offer adequate delay resolution, satisfying the requirement of small enough delay quantization errors \cite{9456029} \cite{tse2005fundamentals}.

To deal with the unknown matrix $\mathbf{C}$, we reformulate (\ref{R_t}) as a sparse model as
\begin{equation}
    \mathbf{R}_{t} = \mathbf{C}^\prime \mathbf{\Pi} \mathbf{B}_{t} + \mathbf{Z}^\prime_{t},\label{R_to}
\end{equation}
where $\mathbf{C}^\prime$ is known and expressed in terms of quantized delay $\mathbf{c}(\ell^\prime/N_d)$ as
    $\mathbf{C}^\prime=\begin{bmatrix}
        \mathbf{c}(1/N_d), \ldots, \mathbf{c}(N_p/N_d) \label{D_o}
    \end{bmatrix}$, and $N_p$ $(L \ll N_p < N_d)$ represents the number of grids used.
$\mathbf{\Pi}$ is an $N_p \times L$ permutation matrix with $L$ nonzero rows, each including a single nonzero element 1, which maps $\mathbf{B}_{t}$ to the corresponding quantized delays. Note that given $\mathbf{R}_{t}$ and $\mathbf{C}^\prime$, we can solve the sparse signal recovery problem in
(\ref{R_to}) to obtain the estimates of delays and  $\mathbf{\Pi} \mathbf{B}_{t}$, based on which the remaining sensing parameters can be efficiently extracted.
\subsection{Sensing Parameter Estimation Algorithm Design}
Traditional spectrum analysis and array signal processing techniques, for instance, MUSIC and ESPRIT require continual observations, which may not always be attainable in PMNs. Therefore, we employ compressive sensing (CS) for estimating sensing parameters from the complicated and fragmented signals, which has an advantage in such problem-solving\cite{8827589}.

We first perform the sparse signal estimation in (\ref{R_to}) to obtain the delay estimates leveraging the CS algorithm UAMP-SBL\cite{9547768}, enabling other sensing parameter estimation and association, where all paths are assumed to have distinct
delays. 
We concatenate $\mathbf{R}_{t}$ corresponding to three consecutive OFDM symbols of the $g$th SSB as 
\begin{equation}
    \tilde{\mathbf{R}}_g = \begin{bmatrix}
        \mathbf{R}_{t^\prime}, \mathbf{R}_{t^\prime + 1}, \mathbf{R}_{t^\prime +2}
    \end{bmatrix} = \mathbf{C}^\prime\mathbf{\Pi}\tilde{\mathbf{B}}_g + \tilde{\mathbf{Z}}^\prime_{g},\label{R_G}
\end{equation}
where $t^\prime = 4g+1$, $\tilde{\mathbf{B}}_g=\begin{bmatrix}
    \mathbf{B}_{t^{\prime}}, \mathbf{B}_{t^{\prime}+1}, \mathbf{B}_{t^{\prime}+2}
\end{bmatrix}$ and $\tilde{\mathbf{Z}}^\prime_{g}=\begin{bmatrix}
    \mathbf{Z}_{t^{\prime}}, \mathbf{Z}_{t^{\prime}+1}, \mathbf{Z}_{t^{\prime}+2}
\end{bmatrix}$.
We can see that the problem in (\ref{R_G}) is a sparse signal estimation one with an $N \times 3M$ observation matrix $\tilde{\mathbf{R}}_g$, an $N\times N_p$ sensing matrix $\mathbf{C}^\prime$ and an $N_p \times 3M$ sparse signal matrix $\mathbf{\Pi}\tilde{\mathbf{B}}_g$, where only $L$ rows  are nonzeros, i.e., the sparse columns in the matrix $\mathbf{\Pi}\tilde{\mathbf{B}}_g$ share a common support, which should be exploited in the estimation of $\mathbf{\Pi}\tilde{\mathbf{B}}_g$. Note that if the beam corresponding to $\mathbf{\Pi}\tilde{\mathbf{B}}_g$ is not pointed to the targets, the power of its nonzero rows can be small so that the targets could be missed. To solve this issue, we concatenate all the matrices $\{\mathbf{\Pi}\tilde{\mathbf{B}}_g\}$ as 
\begin{equation}    \mathbf{A}=\begin{bmatrix}\mathbf{\Pi}\tilde{\mathbf{B}}_1,\mathbf{\Pi}\tilde{\mathbf{B}}_2,\ldots,\mathbf{\Pi}\tilde{\mathbf{B}}_G\end{bmatrix}.\label{B}
\end{equation}
Then we have the composite model
\begin{equation}
\underbrace{\begin{bmatrix}\tilde{\mathbf{R}}_1,\ldots,\tilde{\mathbf{R}}_G\end{bmatrix}}_{\mathbf{R}}=\mathbf{C}^\prime\mathbf{A}+\underbrace{\begin{bmatrix}\tilde{\mathbf{Z}}^\prime_1,\ldots,\tilde{\mathbf{Z}}^\prime_G\end{bmatrix}}_{\mathbf{Z}}.\label{r1rg}
\end{equation}
It is noted that, since all the column vectors in $\mathbf{A}$ share a common support, one can reformulate it as a sparse MMV one. To achieve this, we can force that all the elements in a row of the matrix $\mathbf{\Pi}\tilde{\mathbf{B}}_g$ share a single precision as shown in Algorithm \ref{mmv}.

\begin{algorithm}[t]
    \caption{MMV UAMP-SBL}
    \label{mmv}
    \begin{algorithmic}[1]
    \STATE Unitary transform: $\mathbf{F} = \mathbf{U}^H \mathbf{R} = \boldsymbol{\Phi} \mathbf{A} + \mathbf{U}^H\mathbf{Z}$, where $\boldsymbol{\Phi} = \mathbf{U}^H \mathbf{C'} = \boldsymbol{\Lambda} \mathbf{V}$, and $\mathbf{C'}$ has SVD $\mathbf{C'} = \mathbf{U} \boldsymbol{\Lambda} \mathbf{V}$.
    \STATE Define vector $\boldsymbol{\lambda} = \boldsymbol{\Lambda} \boldsymbol{\Lambda}^H \mathbf{1}$.
    \STATE Initialization: $\forall \upsilon; \tau_{x}^{\upsilon(0)} = 1, \mathbf{\hat{x}}^{\upsilon(0)} = \mathbf{0}, \epsilon' = 0.001,$\\ 
    $\hat{\boldsymbol{\gamma}} = \mathbf{1}, \hat{\beta} = 1, \mathbf{s}^{\upsilon} = \mathbf{0}, t = 0$.
    \REPEAT
    \STATE $\forall \upsilon; \boldsymbol{\tau}_{p}^{\upsilon} = \tau_{x}^{\upsilon(t)} \boldsymbol{\lambda}$
    \STATE $\forall \upsilon; \mathbf{p}^{\upsilon} = \boldsymbol{\Phi} \mathbf{\hat{x}}^{\upsilon(t)} - \boldsymbol{\tau}_{p}^{\upsilon} \cdot \mathbf{s}^{\upsilon}$
    \STATE $\forall \upsilon; \mathbf{v}_{h}^{\upsilon} = \boldsymbol{\tau}_{p}^{\upsilon} ./ (\mathbf{1} + \hat{\beta} \boldsymbol{\tau}_{p}^{\upsilon})$
    \STATE $\forall \upsilon; \mathbf{\hat{h}}^{\upsilon} = (\hat{\beta} \boldsymbol{\tau}_{p}^{\upsilon} \cdot \mathbf{f}^{\upsilon} + \mathbf{p}^{\upsilon}) / (\mathbf{1} + \hat{\beta} \boldsymbol{\tau}_{p}^{\upsilon})$
    \STATE $\hat{\beta} = \Upsilon M / \left(\sum_{\upsilon} (\|\mathbf{f}^{\upsilon} - \mathbf{\hat{h}}^{\upsilon}\|^2 + \mathbf{\mathbf{1}}^T \mathbf{v}_{h}^{\upsilon})\right)$
    \STATE $\forall \upsilon; \boldsymbol{\tau}_{s}^{\upsilon} = \mathbf{1} ./ (\boldsymbol{\tau}_{p}^{\upsilon} + \hat{\beta}^{-\mathbf{1}} \mathbf{\mathbf{1}})$
    \STATE $\forall \upsilon; \mathbf{s}^{\upsilon} = \boldsymbol{\tau}_{s}^{\upsilon} \cdot (\mathbf{f}^{\upsilon} - \mathbf{p}^{\upsilon})$
    \STATE $\forall \upsilon; 1/\tau_{q}^{\upsilon} = (1/N) \lambda^H \boldsymbol{\tau}_{s}^{\upsilon}$
    \STATE $\forall \upsilon; \mathbf{q}^{\upsilon} = \mathbf{\hat{x}}^{\upsilon(t)} + \tau_{q}^{\upsilon} (\boldsymbol{\Phi}^H \mathbf{s}^{\upsilon})$
    \STATE $\forall \upsilon; \tau_{x}^{\upsilon(t+\mathbf{1})} = (\tau_{q}^{\upsilon} / N) \mathbf{\mathbf{1}}^H (\mathbf{1} ./ (\mathbf{1} + \tau_{q}^{\upsilon} \hat{\boldsymbol{\gamma}}))$
    \STATE $\forall \upsilon; \mathbf{\hat{x}}^{\upsilon(t+\mathbf{1})} = \mathbf{q}^{\upsilon} / (\mathbf{1} + \tau_{q}^{\upsilon} \hat{\boldsymbol{\gamma}})$
    \STATE $\hat{\boldsymbol{\gamma}}_{n} = \frac{2\epsilon' + 1 }{ \left(\frac{1}{\Upsilon} \sum_{\upsilon=1}^\Upsilon (|\mathbf{\hat{x}}_{n}^{\upsilon(t+1)}|^2 + \tau_{x}^{\upsilon(t+1)})\right)}, n = 1, \ldots, N$
    \STATE $\epsilon' = \frac{1}{2} \sqrt{\log\left(\frac{1}{N} \sum_{n} \hat{\boldsymbol{\gamma}}_{n}\right) - \frac{1}{N} \sum_{n} \log \hat{\gamma}_{n}}$
    \STATE $t = t + 1$
    \UNTIL{$\frac{1}{\Upsilon} \sum_{\upsilon} (\|\mathbf{\hat{x}}^{\upsilon(t+1)} - \mathbf{\hat{x}}^{\upsilon(t)}\|^2 / \|\mathbf{\hat{x}}^{\upsilon(t+1)}\|^2) \leq \delta_x$ or $t \geq t_{\max}$}
    \end{algorithmic}
\end{algorithm}

After performing sparse signal estimation, we can identify the $L$ rows corresponding to the $L$ paths in $\mathbf{A}$. Then we remove the rows with elements close to zero in $\mathbf{A}$, leading to
\begin{equation}
    \mathbf{A}^\prime = \mathbf{\Pi}^\prime \begin{bmatrix} \tilde{\mathbf{B}}_1, \ldots, \tilde{\mathbf{B}}_G \end{bmatrix},
\end{equation}
where $\mathbf{\Pi}^\prime$ is an $L \times L$ matrix used for removing relevant rows. Note that the multipath index $l=1, \ldots, L$ is not associated with the row index $i = 1,\ldots, L$ in $\mathbf{A}^\prime$, so the delay estimation can be given in terms of $i$ as
\begin{equation}
    \hat{\tau}_i = \frac{i'}{N_d \Delta f},\label{taur}
\end{equation}
where $i^\prime$ is the row index in $\mathbf{A}$ that corresponds to the $i$th row in $\mathbf{A}^\prime$.

To identify the beams aligned with multipath, we introduce $\mathbf{d}^T_{i,g} = [\mathbf{b}^T_{i,t^\prime}, \mathbf{b}^T_{i,t^\prime+1}, \mathbf{b}^T_{i,t^\prime+2}]$ and represent $\mathbf{A}^\prime$ as
\begin{equation}
    \begin{aligned}
        \mathbf{A}^\prime=\begin{bmatrix}    \mathbf{d}^T_{1,1} & \mathbf{d}^T_{1,2} & \ldots & \mathbf{d}^T_{1,G} \\
                   \vdots             & \vdots             & \vdots & \vdots             \\
                   \mathbf{d}^T_{L,1} & \mathbf{d}^T_{L,2} & \ldots & \mathbf{d}^T_{L,G}\end{bmatrix}, \label{B^p}
    \end{aligned}
\end{equation}
where each row of $\mathbf{A}^\prime$ corresponds to one path. If the $g$th SSB beam is aligned with the $i$th path, $\mathbf{d}^T_{i,g}$ has higher power than that of other SSBs. To improve the estimation accuracy compared with that using any $\mathbf{d}^T_{i,g}$, we need to determine the optimal ${g}_i$ with the largest $\Vert \mathbf{d}^T_{i,g} \Vert_2$ in the $i$th row of $\mathbf{A}^\prime$, i.e.,
\begin{equation}
    \hat{g}_i = \operatorname*{argmax}_{g=1, \ldots, G} \Vert \mathbf{d}^T_{i,g} \Vert_2. \label{g_i}
\end{equation}

From $\mathbf{b}^{T}_{i,t^{\prime}_i}=a_ib_{i}e^{j2\pi t^{\prime}_if_{D,i}T_s}\mathbf{w}^{T}_{q_i,t^{\prime}_i}\mathbf{a}(M,\theta_{i})\mathbf{a}^T(M,\theta_{i})$ where $t^{\prime}_i=4g_i+1$, we can obtain $\hat{f}_{D,i}$, $\hat{\theta}_i$ and $\left|\hat{b}_i\right|^2$ by
\begin{equation}
    \hat{f}_{D,i} = \frac{1}{-2\pi T_s} \angle \left(\sum_{j=0}^{1}\mathbf{b}^T_{i,t^{\prime}_i+j} \mathbf{b}^{\ast}_{i,t^{\prime}_i+j+1}\right),\label{fd}
\end{equation}
\begin{equation}
    \begin{aligned}
        \text{sin}(\hat{\theta}_{i})&=\frac{1}{-\pi}\!\angle\!\left(\sum_{j=0}^{2}\!\sum_{n=1}^{M-1}\!\mathbf{b}^{T}_{i,t^{\prime}_i +j}(n)\mathbf{b}^{H}_{i,t^{\prime}_i+j}(n\!+\!1)\right),
    \end{aligned} \label{phi}
\end{equation}
and
\begin{equation}
\left|\hat{b}_i\right|^2\! =\! (2M)^{-1}\left| \hat{a}_i \mathbf{w}^H_{t_i^\prime+1} \mathbf{a}^*(M, \hat{\theta}_i)\right|^{-2}\left| \sum_{j=0}^{1}  \mathbf{b}_{i,t_i^\prime+1}^T \mathbf{b}_{i,t_i^\prime+2j}^* \right| .\quad \label{br}
\end{equation}

Now we have estimated and associated all sensing parameters of $\{\tau_i, f_{D,i}, \theta_i, b_i\}$ by the index $i$. The procedure of sensing parameter estimation with direct clutter mitigation is summarized in Algorithm \ref{alg:estimation}.
\subsection{Complexity Analysis}

In the proposed method, direct clutter mitigation is first performed, which only involves simple operations in (\ref{ynttilde}) with complexity $O(PNM)$. In contrast, the complexity of GMM-EM-CE is \(O(KNN_mL_cM^3)\), where $K$ is the number of iterations, $N_m$ is the number of samples used, and $L_c$ is the number of the multivariate Gaussian mixture components \cite{9203851}. Hence, the complexity of the proposed method is far lower than that of GMM-EM-CE. In terms of sensing parameter estimation, the proposed method requires $ \mathcal {O} (NN_pG)$ per iteration for delay estimation, and $ \mathcal {O} (LM)$ for the remaining parameters estimation in (\ref{fd})-(\ref{br}). Note that the SVD operation in the proposed method can be carried out offline. Compared to the indirect sensing parameter estimation method proposed in \cite{8827589} that has the computational complexity of $\mathcal {O} (NN_p^2)$ per iteration, our proposed approach has lower complexity as typically $G \ll N_p$. 



\begin{algorithm}[t]
    \caption{Sensing Parameter Estimation with Direct Clutter Mitigation}
    \label{alg:estimation}
    \begin{algorithmic}[1]
        \REQUIRE signals $\mathbf{{y}}_{n,t}$, $s_{n,t}$, threshold $\gamma$.
        \STATE Compute $\mathbf{\tilde{y}}_{n,t}$ by (\ref{ynttilde}).
        \FOR{$t = 1$ to $4G+3$}
            \FOR{$ n = 1$ to $N$} 
                \STATE Compute $\mathbf{r}_{n, t}$ by (\ref{rnt}).
            \ENDFOR
            \STATE Construct $\mathbf{R}_t$ by (\ref{R_t}).
        \ENDFOR
        \FOR{$g=1$ to $G$}
            \STATE Construct $\tilde{\mathbf{R}}_g$ by (\ref{R_G}).
            
        \ENDFOR
        \STATE Estimate $\mathbf{A}$ from (\ref{r1rg}) using Algorithm \ref{mmv}.
        \STATE Obtain $\mathbf{A}^\prime$ by removing rows whose 2-norms are less than $\gamma$ in $\mathbf{A}$.
        \FOR{$i = 1$ to $L$}
        \STATE Estimate $\hat{\tau}_i$, $\hat{f}_{D,i}$, $\hat{\theta}_i$ and  $\hat{b}_i$ by (\ref{taur}) and (\ref{fd})-(\ref{br}), respectively.  
        \ENDFOR
    \end{algorithmic}
\end{algorithm}

\section{Simulation Results}

Consider a PMN with a single RRU equipped with four antennas, and its transmission power is set to 30 dBm. The carrier frequency and signal bandwidth are set to 2.35 GHz and 100 MHz, respectively. The receiver thermal noise power is calculated by $N_0 + 10log(10^8) = -94$ dBm with a power spectral density of $N_0 = -174$ dBm/Hz. SSB signals with $G = 4$ are used for sensing, which occupy 240 subcarriers and 12 symbols across the time-frequency domain. The radar cross section of the targets is assumed to be 1 m² in the simulations.

Multipath signals are randomly generated in cluster, mimicking reflected/scattered signals from objects. The large-scale path loss model is employed with a pathloss factor 4. The number of multipath, angle span, target distance and Doppler frequency are assumed to be uniformly distributed in the range of [15, 25], [0, 45] degrees, [0, 60] m and [0, 600] Hz, respectively, where clutter signals are included with near-zero Doppler-frequencies. Additional offsets between objects and RRU include angle offsets of [-75, 75] degrees, distance offsets of [60, 120] m and moving speed offsets of [-40, 40] m/s. Delays are quantized to 10 ns on a grid of $N_d = 512$, which is equivalent to a distance quantization of 3 m. Based on pathloss factor and multipath propagation distances, the received signal-to-noise ratio (SNR) can be as low as 0 dB when estimating the sensing parameters.

In simulations, we adopt $P=2$ and recursive filtering to enhance the stability and accuracy of the sensed parameters over time. It iteratively refines the parameter estimates by incorporating new estimates, i.e., $\mathbf{\hat{q}}(i) = \alpha \mathbf{\hat{q}}(i-1) + (1 - \alpha) \mathbf{q}(i)$, where $\mathbf{\hat{q}}(i)$ and $\mathbf{q}(i)$ denote the filtered and new parameter estimate vector for the $i$th SSB burst set respectively, and $\alpha = 0.9$ is the forgetting factor. This smooths out the noise and reduce the impact of outlier measurements, which is particularly beneficial in dynamic environments. 

\begin{figure}[t]
    \centering
    \includegraphics[width=3.3in]{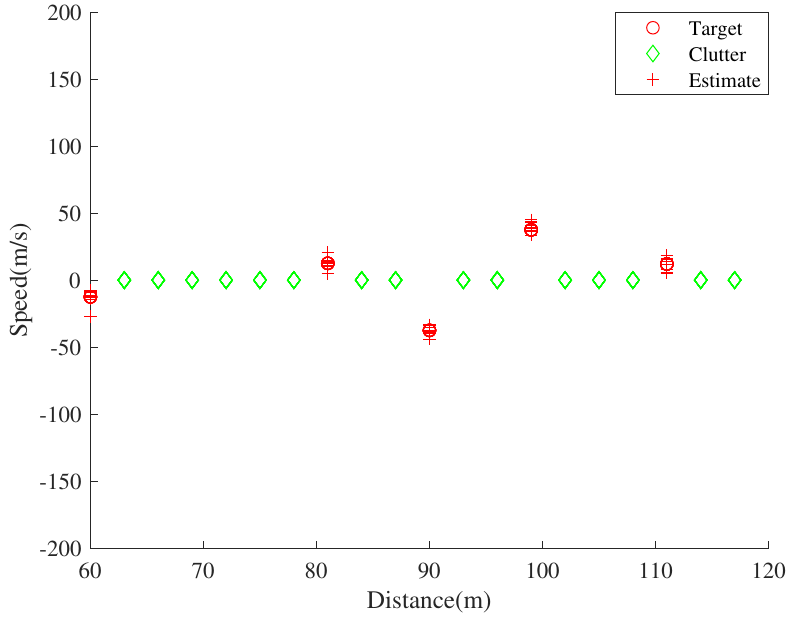}
    \caption{ Ten trials for speed-distance estimation.}\label{dop}
\end{figure}

\begin{figure}[htpb]
    \centering
    \includegraphics[width=3.3in]{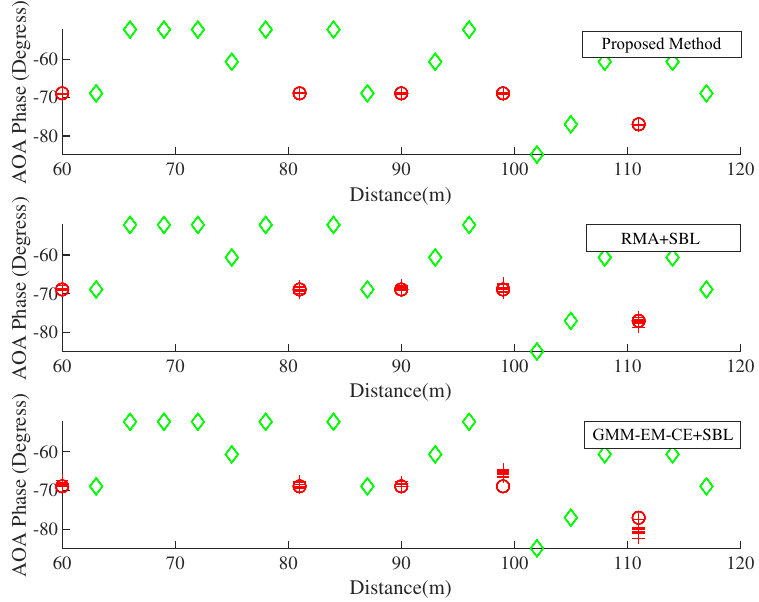}
    \caption{ Ten trials for AoA-distance estimation.}\label{result}
\end{figure}

Fig.~\ref{dop} shows the speed estimation performance versus target distance using the proposed method, highlighting that clutters with near-zero velocity can be effectively suppressed while targets are well preserved and the corresponding speeds are accurately estimated.

Fig.~\ref{result} compares the AoA estimation results in terms of \( \pi \sin(\theta_l) \) versus target distance for the proposed one, RMA + SBL \cite{8827589} and GMM-EM-CE + SBL \cite{9203851}. It can be seen from the results that regardless of distance, the proposed one has more reliable and accurate AoA estimates than RMA + SBL and GMM-EM-CE + SBL. This is attributed to the robustness and high performance of the UAMP algorithm. 
Additionally, the structured variational inference used in UAMP-SBL efficiently exploits a sparse prior to recover sparse signals. The GMM differentiates target signals and clutter using probabilistic characteristics over a time interval of 20$T_s$ with 10 iterations and 64 samples. The RMA leverages the stability of sensing parameters over short intervals and a recursive averaging process with a forgetting factor of 0.99 over 64 iterations. 
\begin{figure}[t]
    \centering
    \includegraphics[width=3.3in]{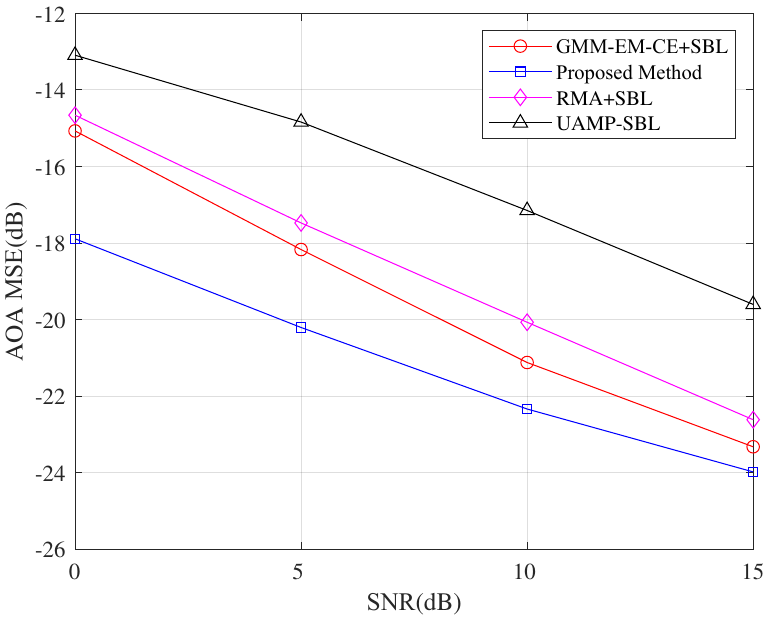}
    \caption{Comparison of various methods in terms of AOA estimation MSE}\label{compare}
\end{figure}

Fig.~\ref{compare} compares the mean square errors (MSEs) in AoA estimation versus SNR. It is shown that at the MSE of -20dB, the proposed one outperforms GMM-EM-CE + SBL and RMA + SBL with SNR gains of 3dB and 5dB, respectively. In addition, the proposed one performs much better than the UAMP-SBL one that does not perform clutter suppression, demonstrating the necessity and effectiveness of the clutter suppression inbuilt in the proposed method. 

\section{Conclusion}
This paper has investigated the issue of sensing parameter estimation in the presence of clutter in PMNs. We propose a novel low-complexity method by integrating multipulse canceller based clutter suppression mechanism into sensing parameter estimation leveraging UAMP, where parameter association is also performed. In particular, the beam scanning property of 5G NR SSB signals is exploited. 
Simulation results demonstrate the advantages of the proposed method in sensing performance and complexity compared to the existing ones.

\bibliographystyle{IEEEtran}
\bibliography{reference.bib}

\end{document}